\documentstyle[aps,eqsecnum,preprint,tighten,floats,epsf,rotate]{revtex}

\newbox\rotbox

\begin{document}
\draft


\preprint{\vbox{\it Submitted to Physical Review C
                                        \null\hfill\rm TRI-PP-96-8\\
                                         \null\hfill\rm nucl-th/9604018 }}
%
\title{MODIFIED QUARK-MESON COUPLING MODEL\\ FOR NUCLEAR MATTER}
\author{Xuemin Jin\thanks{%
Address after September 1, 1996: Center for Theoretical Physics,
Laboratory for Nuclear Science and Department of Physics,
Massachusetts Institute of Technology, Cambridge, 
Massachusetts 02139}
 and B.K.~Jennings}
\address{TRIUMF, 4004 Wesbrook Mall, Vancouver, \\British Columbia, 
         Canada V6T 2A3} 
%
%
\maketitle
\begin{abstract}
The quark-meson coupling model for nuclear matter, which describes nuclear 
matter as non-overlapping MIT bags bound by the self-consistent exchange of 
scalar and vector mesons, is modified by introducing medium modification 
of the bag constant. We model the density dependence of the bag constant in
two different ways: one invokes a direct coupling of the bag constant to the
scalar meson field, and the other relates the bag constant to the in-medium nucleon 
mass. Both models feature a decreasing bag constant with increasing density.
We find that when the bag constant is significantly 
reduced in nuclear medium with respect to its free-space value, large 
canceling isoscalar Lorentz scalar and vector potentials for the nucleon 
in nuclear matter emerge naturally. Such potentials are comparable to those 
suggested by relativistic nuclear phenomenology and finite-density QCD sum
rules. This suggests that the reduction of bag constant in nuclear medium may 
play an important role in low- and medium-energy nuclear physics. 
\end{abstract}
\pacs{PACS numbers: 24.85+p; 21.65+f; 12.39.Ba; 12.38.Lg}

\narrowtext
%
\section{Introduction}
\label{intro}

Ultimately, the physics of nuclear matter and finite nuclei is an
exercise in applied quantum chromodynamics (QCD), which governs the
underlying strong interactions of quarks and gluons.  In reality,
however, knowledge of QCD has had very little impact, to date, on the
study of low- and medium-energy nuclear phenomena. The reason is that
QCD is intractable at the nuclear physics energy scales due to the
nonperturbative features of QCD. A reasonable consensus is that the
relevant degrees of freedom for low energy QCD are hadrons instead of
quarks and gluons.

While the description of nuclear phenomena has been efficiently
formulated using the hadronic degrees of freedom, new challenges arise
from the observed small but interesting corrections to the standard
hadronic picture such as the EMC effect which reveals the medium
modification of the internal structure of nucleon \cite{emc}. To
address these new challenges, it is necessary to build theories that
incorporate quark-gluon degrees of freedom, yet respect the
established theories based on hadronic degrees of freedom.

A few years ago, Guichon\cite{guichon88} proposed a quark-meson
coupling (QMC) model to investigate the direct ``quark effects'' in
nuclei. This model describes nuclear matter as non-overlapping MIT
bags interacting through the self-consistent exchange of mesons in the
mean-field approximation. This simple model was refined later by
including nucleon Fermi motion and the center-of-mass corrections to
the bag energy \cite{fleck90} and applied to variety of
problems\cite{saito94,saito94a,saito92,song95}.  Recently, the QMC
model has been applied to finite nuclei \cite{guichon95,blunden96}.
(There have been several works that also discuss the quark effects in
nuclei, based on other effective models for the nucleon
\cite{banerjee92}).

Although it provides a simple and attractive framework to incorporate
the quark structure of the nucleon in the study of nuclear phenomena,
the QMC model has a serious short-coming.  It predicts much smaller
scalar and vector potentials for the nucleon than obtained in
relativistic nuclear phenomenology.  Unless there is a large isoscalar
anomalous coupling (ruled out by other considerations) this implies a
much smaller nucleon spin-orbit force in finite nuclei.  To lowest
order in the nucleon velocity and potential depth the nucleon
spin-orbit potential can be obtained in a model independent way from
the strengths of the scalar and vector potentials. The spin-orbit
potential from the QMC model is too weak to successfully explain
spin-orbit splittings in finite nuclei and the spin-observables in
nucleon-nucleus scattering.

Relativistic nuclear phenomenology is a general approach based on
nucleons and mesons and has gained tremendous credibility during last
twenty years. In this framework, the nucleons in nuclear environment
are treated as point-like Dirac particles interacting with large
canceling isoscalar Lorentz scalar and vector potentials. This
approach has been successful in describing the spin-observables of
nucleon-nucleus scattering in the context of relativistic optical
potentials\cite{hama90,wallace87}.  Moreover, such potentials can be
derived from the relativistic impulse approximation
\cite{wallace87}. The relativistic field-theoretical models based on
nucleons and mesons, QHD, also feature Dirac nucleons interacting
through the exchange of scalar and vector mesons \cite{serot86}. QHD,
at the mean-field level, has proven to be a powerful tool for
describing the bulk properties of nuclear matter and spin-orbit
splittings of finite nuclei \cite{serot86}. It is known that the large
and canceling scalar and vector potentials are central to the success
of the relativistic nuclear phenomenology. Recent progress in
understanding the origin of these large potentials for propagating
nucleons in nuclear matter has been made via the analysis of the
finite-density QCD sum rules \cite{cohen95}.

In a recent paper \cite{jin96}, the present authors have pointed out
that the resulting small nucleon potentials in the QMC model stem from
the assumption of fixing the bag constant at its free-space value, and
that this assumption is questionable. We then included a density
dependent bag constant and found that when the bag constant drops
significantly in nuclear matter relative to its free-space value, the
large potentials for nucleons in nuclear matter, as seen in the
relativistic nuclear phenomenology and finite-density QCD sum rules,
can be recovered. This suggests that the reduction of bag constant 
in nuclear matter relative to its free-space value may be essential for 
the successes of relativistic nuclear phenomenology and thus may play 
an important role in low- and medium-energy nuclear physics. In the 
present paper, we present further details and model the density dependence 
of the bag constant in two different ways: one invokes a direct coupling 
of the bag constant to the scalar meson field, and the other relates the 
bag constant to the in-medium nucleon mass.

This paper is organized as follows: In Section~\ref{qmc-model} we
sketch the QMC model for nuclear matter. We then modify the QMC model
by introducing a density dependent bag constant in nuclear matter in
Section~\ref{chiral}. The results are presented in
Section~\ref{result}. Further discussions are given in
Section~\ref{discussion}.  Section~\ref{conclusion} is a summary.

%
\section{The quark-meson coupling model for nuclear matter}
\label{qmc-model}

In this section, we give a brief introduction to the quark-meson
coupling model for nuclear matter. The reader is referred to
Refs. \cite{guichon88,fleck90,saito94,saito94a} for further details
and justifications for using a simple bag model for the in-medium
nucleon.

In the QMC model, the nucleon in nuclear medium is assumed to be a
static spherical MIT bag in which quarks interact with the scalar and
vector fields, $\overline{\sigma}$ and $\overline{\omega}$, and these
fields are treated as classical fields in the mean field
approximation. (Here we only consider up and down quarks.)  The quark
field, $\psi_q(x)$, inside the bag then satisfies the equation of
motion: 
\begin{equation}
\left[i\,\rlap{/}\partial-(m_q^0-g_\sigma^q\, \overline{\sigma})
-g_\omega^q\, \overline{\omega}\,\gamma^0\right]\,\psi_q(x)=0\ ,
\label{eq-motion}
\end{equation}
where $m_q^0$ is the current quark mass, and $g_\sigma^q$ and
$g_\omega^q$ denote the quark-meson coupling constants.  We will
neglect isospin breaking and take $m_q^0=(m_u^0+m_d^0)/2$ hereafter.
The normalized ground state for a quark in the bag is given
by\cite{guichon88,fleck90,saito94}
\begin{equation}
\psi_q(t,{\bf r})={\cal N}\, e^{-i\epsilon_q t/R}
\left(
\begin{array}{c}
j_0(xr/R)
\\*[7.2pt]
i\,\beta_q\, {\bf \sigma\cdot \hat{r}}\, j_1(xr/R)
\end{array}
\right)\, {\chi_q\over \sqrt{4\pi}}\ ,
\label{wave-function}
\end{equation}
where 
\begin{eqnarray}
& &\epsilon_q=\Omega_q + g_\omega^q\, \overline{\omega}\, R\ ,\hspace*{1cm}
\beta_q=\sqrt{{\Omega_q-R\, m_q^*\over 
\Omega_q\, +R\, m_q^*}}\ ,
\\*[7.2pt]
& &{\cal N}^{-2}=2\, R^3 \, j_0^2(x)\left[
\Omega_q(\Omega_q-1)+R\, m_q^*/2\right]/x^2\ ,
\end{eqnarray}
with $\Omega_q\equiv \sqrt{x^2+(R\, m_q^*)^2}$, 
$m_q^*=m_q^0-g_\sigma^q\, \overline{\sigma}$, $R$ the bag radius, 
and  $\chi_q$ the quark spinor. The $x$ value is determined by the 
boundary condition at the bag surface
\begin{equation}
j_0(x)=\beta_q\, j_1(x)\ .
\label{bun-con}
\end{equation}

The energy of a static bag consisting of three ground state quarks 
can be expressed as
\begin{equation}
E_{\rm bag}=3\, {\Omega_q\over R}-{Z\over R}
+{4\over 3}\,  \pi \, R^3\,  B\ ,
\label{ebag}
\end{equation}
where $Z$ is a parameter which accounts for zero-point motion
and $B$ is the bag constant. In the discussions to follow, we 
use $R_0$, $B_0$ and $Z_0$ to denote the corresponding bag parameters
for the free nucleon. After the corrections of spurious 
center-of-mass motion in the bag, the effective mass of a nucleon bag at rest
is taken to be\cite{fleck90,saito94}
\begin{equation}
M_N^*=\sqrt{E_{\rm bag}^2-\langle p_{\rm cm}^2\rangle}\ ,
\label{eff-mn}
\end{equation}
where $\langle p_{\rm cm}^2\rangle=\sum_q \langle p_q^2\rangle$ and
$\langle p_q^2\rangle$ is the expectation value of the quark momentum
squared, $(x/R)^2$. 

The equilibrium condition for the bag is obtained by 
minimizing the effective mass $M_N^*$ with respect to the bag radius
\begin{equation}
{\partial M_N^*\over \partial R} = 0\ .
\label{balance}
\end{equation}
In free space, one may fix $M_N$ at 
its experimental value $939$ MeV and use the equilibrium condition
to determine the bag parameters. For several choices of bag radius,
$R_0 = 0.6, 0.8, 1.0$ fm, the results for $B_0^{1/4}$ and $Z_0$
are $188.1, 157.5, 136.3$ MeV and $2.030,1.628,1.153$, respectively.

The total energy per nucleon at finite density $\rho_N$, including the
Fermi motion of the nucleons, can be 
written as\cite{saito94}
\begin{eqnarray}
E_{\rm tot}& = &{\gamma\over (2\pi)^3\, \rho_N} 
\int^{k_F}\, d^3 k \sqrt{M_N^{* 2}+{\bf k}^2}
+{g_\omega^2\over 2 m_\omega^2}\,\rho_N
+{m_\sigma^2\over 2\,\rho_N}\overline{\sigma}^2\ ,
\label{etot}
\end{eqnarray}
where $\gamma$ is the spin-isospin degeneracy, and $\gamma=4$ for
symmetric nuclear matter and $\gamma=2$ for neutron matter.\footnote{%
Here we only consider symmetric nuclear matter and neutron matter. The
generalization to a general asymmetric nuclear matter is straightforward
(see, for example, Ref. \cite{saito94}).}
Note that the mean field $\overline{\omega}$ created
by uniformly distributed nucleons is determined by baryon number 
conservation to be\cite{guichon88,fleck90,saito94}
\begin{equation}
\overline{\omega}={3\, q^q_\omega\, \rho_N\over m_\omega^2}
= {g_\omega\,\rho_N\over m_\omega^2}\ ,
\label{vec-field}
\end{equation}
where $g_\omega\equiv 3 g^q_\omega$.
The scalar mean field is determined by the thermodynamic condition
\begin{equation}
\left({\partial\, E_{\rm tot}\over 
\partial\, \overline{\sigma} }\right)_{R,\rho_N} = 0\ .
\label{thermal}
\end{equation}
If one {\it assumes}
\begin{equation}
B=B_0
\label{qmc-ass}
\end{equation}
and $Z=Z_0$, Eq.~(\ref{thermal})
yields the self-consistency condition 
\begin{equation}
\overline{\sigma} = {g_\sigma \over m_\sigma^2}\, 
C (\overline{\sigma})
{\gamma\over (2\pi)^3} 
\int^{k_F}\, d^3 k\, {M_N^*\over 
\sqrt{M_N^{* 2}+{\bf k}^2}} \ ,
\label{scc}
\end{equation}
with
\begin{equation}
g_\sigma\, C(\overline{\sigma}) = g_\sigma\, {E_{\rm bag}\over M_N^*}
\Biggl[\left(1-{\Omega_q\over E_{\rm bag}\, R}\right)\,
S(\overline{\sigma}) +
{m_q^*\over E_{\rm bag}}\Biggr] \ ,
\label{tc}
\end{equation}
where $g_\sigma\equiv 3\, g^q_\sigma$ and 
\begin{equation}
S(\overline{\sigma}) = 
{\Omega_q/2+R\,m_q^*\,(\Omega_q-1)\over
\Omega_q\,(\Omega_q-1)+R\,m^*_q/2}\ .
\end{equation}
The two coupling constants $g_\sigma$ and $g_\omega$ can be
chosen to fit the nuclear matter binding energy at the saturation
density. For a given density, Eqs.~(\ref{bun-con}), (\ref{balance}),
and (\ref{scc}) form a set of equations for calculating $x$, $R$, and
$\overline{\sigma}$.

%
\section{Modified quark-meson coupling model}
\label{chiral}

In this section, we modify the QMC model by introducing a density
dependent bag constant. We propose two models for the modification of
the bag constant, featuring a decreasing bag constant with increasing 
nuclear matter density. In principle, the parameter $Z$ may also be modified
in the nuclear medium. However, it is unclear how $Z$ changes with the density.
Here we assume that the medium modification of $Z$ is small at low and moderate 
densities and take $Z=Z_0$.\footnote{Recently, Blunden and Miller \cite{blunden96} have
considered a density dependent $Z$. However, it is found that for
reasonable parameter ranges changing $Z$ has little effect and tend to
make the model worse.}

%
\subsection{Direct coupling model}

The bag constant in the MIT bag model contributes $\sim 200-300$ MeV
to the nucleon energy and provides the necessary pressure to confine
the quarks.  Thus, the bag constant is an inseparable ingredient of
the bag picture of a nucleon. When a nucleon bag is put into the
nuclear medium, the bag as a whole reacts to the environment. As a
result, the bag constant may be modified.  There is little doubt that
at sufficiently high densities, the bag constant is eventually melted
away and quarks and gluons become the appropriate degrees of freedom.
Therefore, It is reasonable to believe that the bag constant is modified and
decreases as density increases. This physics is obviously bypassed in
the QMC model by the assumption of $B=B_0$.

To reflect this physics, we modify the QMC model by introducing
a direct coupling between the bag constant and the scalar mean field
\begin{equation}
{B\over B_0} = \left[ 1
- g_\sigma^B\, {4 \over \delta} {\overline{\sigma}\over M_N} \right]^\delta\ ,
\label{an-dir}
\end{equation}
where $g_\sigma^B$ and $\delta$ are real positive parameters and the
introduction of $M_N$ is based on the consideration of dimension.
(The case, $\delta = 1$, was also considered by Blunden and Miller
\cite{blunden96}.)  Note that $g_\sigma^B$ differs from $g_\sigma^q$
(or $g_\sigma$).  When $g_\sigma^B = 0$, the usual QMC model is
recovered.  

This direct coupling can be partially motivated from
considering a non-topological soliton model for the nucleon where a
scalar soliton field provides the confinement of the quarks. Roughly
speaking, the bag constant in the MIT bag model mimics the effect of
the scalar soliton field in the soliton model. Now, when a nucleon
soliton is put into nuclear environment, the scalar soliton field will
interact with the scalar mean field (see, for example, 
Refs.~\cite{cahill96,banerjee92}). Therefore, it is reasonable to
couple the bag constant directly to the scalar mean field.

The factor $C(\overline{\sigma})$ of (\ref{tc}), appearing 
in the self-consistency condition Eq.~(\ref{scc}), then becomes 
\begin{equation}
g_\sigma C(\overline{\sigma}) =
g_\sigma {E_{\rm bag}\over M_N^*}
\Biggl[\left(1-{\Omega_q\over E_{\rm bag}\, R}\right)\,
S(\overline{\sigma}) +
{m_q^*\over E_{\rm bag}}\Biggr] +
g_\sigma^B\,
{E_{\rm bag}\over M_N^*}\,
{16\over 3}\,\pi\, R^3\,{B\over M_N}\,
\Biggl[1-{4 \over \delta}\;{g_\sigma^B \overline{\sigma}\over M_N}
\Biggr]^{-1}\ .
\label{tc-d}
\end{equation}
The other equations are not affected. In the limit of $\delta \rightarrow
\infty$, Eq.~(\ref{an-dir}) reduces to an exponential form with a
single parameter $g_\sigma^B$
\begin{equation}
{B\over B_0} = e^{- 4\, g^B_\sigma\, \overline{\sigma}/M_N}\ .
\label{dir-limit}
\end{equation} 
In the limit of zero current quark mass (i.e., $m_q^0 = 0$) and $g_\sigma=0$,
the nucleon mass scales like
$B^{1/4}$ from dimensional arguments (see also Appendix A). Then from 
Eq.~(\ref{an-dir}) we get
%
$ M_N^*/M_N  = \left( B/B_0  \right)^{1/4} = \left[ 1
- g_\sigma^B\, {4 \over \delta} {\overline{\sigma}\over M_N} 
\right]^{\delta/4}$.
%
We observe that the linear $\sigma$-nucleon coupling is just
$g_\sigma^B$ while $\delta$ controls the non-linearities. For
$\delta=4$, the non-linearities vanish and, as discussed in the next
section, we recover QHD-I but with a density dependent bag radius.
%
\subsection{Scaling model}

In the previous paper \cite{jin96}, we have considered a scaling
model, which relates the in-medium bag constant to the in-medium
nucleon mass directly
\begin{equation}
{B\over B_0} = \left[ M_N^*\over M_N \right]^\kappa\ ,
\label{an-br}
\end{equation}
where $\kappa$ is a real positive parameter and $\kappa=0$ corresponds 
to the usual QMC model. The factor $C(\overline{\sigma})$, in this case, 
is given by
\begin{equation}
g_\sigma\, C(\overline{\sigma}) = g_\sigma
{E_{\rm bag}\over M_N^*}
\Biggl[\left(1-{\Omega_q\over E_{\rm bag}\, R}\right)\,
S(\overline{\sigma}) +
{m_q^*\over E_{\rm bag}}\Biggr]
\Biggl[1-\kappa\, {E_{\rm bag}\over M_N^{* 2}}
{4\over 3}\,\pi\,R^3\, B\,
\Biggr]^{-1}\ .
\label{tc-k}
\end{equation}
Note that in this model, the effective nucleon mass $M_N^*$ and the bag
constant $B$ are determined self-consistently by combining
Eqs.~(\ref{ebag}), (\ref{eff-mn}), and (\ref{an-br}).

One notices that both Eqs.~(\ref{an-dir}) and ({\ref{an-br}) give rise
to a reduction of the bag constant in nuclear medium relative to its
free-space value. While the scaling model is characterized by a single
free parameter $\kappa$, it leads to a complicated and implicit 
relation between the bag constant and the scalar mean field. On the other
hand, the direct coupling model features a straightforward coupling
between the bag constant and the scalar mean field, which, however, 
introduces two free parameters, $g_\sigma^B$ and $\delta$.

\section{Results} \label{result}

In this section, we present numerical results. The two models for the
in-medium bag constant discussed in previous section will be
considered. The current quark masses are taken to be $m_u = m_d =0$
for simplicity. Inclusion of small current quark masses only leads to
numerically small refinement of present results.

Let us start from the direct coupling model. For a given
value of $g^q_\sigma$, we adjust the coupling constants $g^B_\sigma$
and $g_\omega$ to reproduce the nuclear matter binding energy ($-16$ MeV) 
at the saturation density ($\rho_{\rm N}^0=$0.17 fm$^{-3}$). 
The resulting coupling constants and nuclear matter results
are given in Table~\ref{tab-1} for various $\delta$ values with $R_0 =
0.6$ fm. For the special case, $g_\sigma^q = 0$ and $\delta = 4$, the
present model leads to exactly the same nuclear matter results as
obtained in QHD-I (see first row of Table~\ref{tab-1}). This is also
shown analytically in Appendix A.

\begin{table}
\caption{Coupling constants and nuclear matter results as obtained from 
the direct coupling model. The free space bag radius is fixed at $R_0 = 
0.6$ fm. Here the nuclear matter compressibility, $K_V^{-1}$, is given 
in unit of MeV, $r_m^*$ and $r_m$ denote the quark root-mean-square radius 
in nuclear matter and in free space, respectively, and 
$U_{\rm v}\equiv g_\omega\, \overline{\omega}$ is the vector mean field. 
The mass parameters are taken to be $m_q=0$, $m_\sigma=550$ MeV, and
$m_\omega=783$ MeV.}
\label{tab-1}
\begin{tabular}{cccccccccccc}
$g_\sigma^q$ &$\delta$  &$(g_\sigma^B)^2/4\pi$  &$g_\omega^2/4\pi$ 
&$M_N^*/M_N$  &$U_{\rm v}/M_N$
&$K_V^{-1}$ &$B/B_0$ &$x/x_0$ &$R/R_0$ &$r_m^*/r_m$\\
\tableline
\tableline
0    & 4   & 8.45  & 12.84 & 0.55 & 0.37 & 540  & 0.09  & 1.0 & 1.83 & 1.83 \\
     & 8   & 5.68  & 6.46  & 0.75 & 0.18 & 313  & 0.31  & 1.0 & 1.34 & 1.34 \\
     & 12  & 5.40  & 5.68  & 0.77 & 0.16 & 295  & 0.35  & 1.0 & 1.30 & 1.30 \\
     & 13  & 5.28  & 5.57  & 0.77 & 0.16 & 293  & 0.36  & 1.0 & 1.29 & 1.29 \\
 &$\infty$ & 4.95  & 4.62  & 0.80 & 0.13 & 270  & 0.41  & 1.0 & 1.25 & 1.25 \\
\tableline
1.0  & 4   & 5.69  & 10.84 & 0.61 & 0.32 & 490  & 0.19  & 0.97& 1.51 & 1.52 \\
     & 8   & 4.20  & 6.78  & 0.74 & 0.19 & 333  & 0.36  & 0.97& 1.28 & 1.29 \\
     & 12  & 3.96  & 6.14  & 0.76 & 0.18 & 315  & 0.39  & 0.97& 1.25 & 1.26 \\
 &$\infty$ & 3.69  & 5.24  & 0.78 & 0.15 & 289  & 0.45  & 0.98& 1.22 & 1.22 \\
\tableline
2.0  &3.6  & 3.16  & 8.03  & 0.70 & 0.23 & 431  & 0.36  & 0.93& 1.27 & 1.29 \\
     & 4   & 2.99  & 7.42  & 0.72 & 0.21 & 398  & 0.39  & 0.94& 1.25 & 1.27 \\
     & 8   & 2.54  & 5.81  & 0.77 & 0.17 & 336  & 0.48  & 0.95& 1.18 & 1.20 \\
     & 12  & 2.43  & 5.48  & 0.76 & 0.16 & 324  & 0.50  & 0.95& 1.17 & 1.19 \\
 &$\infty$ & 2.30  & 4.96  & 0.79 & 0.14 & 305  & 0.54  & 0.95& 1.15 & 1.17 \\
\tableline
5.309 &$\text{--}$ & $0.0$ & 1.56 & 0.89 & 0.04 & 223  & 1.0  & 0.93 &
 0.98 & 1.0  \\
\end{tabular}
\end{table}
%

The most important feature is that the reduction of $B$ relative to $B_0$ 
leads to the decrease of $M_N^*/M_N$ and the increase of $U_{\rm v}/M_N$ 
relative to their values in the simple QMC model. In the usual QMC model,
the required vector coupling is very small.  This, 
in Refs.~\cite{fleck90,saito94}, is attributed to the repulsion provided 
by the center-of-mass corrections to the bag energy. In our modified 
QMC model, the reduction of the bag constant in nuclear medium provides a
new source of attraction as it effectively reduces $M_N^*$. Consequently, 
additional vector field strength is required to obtain the correct 
saturation properties of nuclear matter.

The above physics is clearly reflected in Table~\ref{tab-1}. 
It can be seen from Table \ref{tab-1} that when the bag constant is
reduced significantly in nuclear matter relative to its free-space
value, the resulting magnitudes for $M_N^* - M_N$ and $U_{\rm v} \equiv 
g_\omega \overline{\omega}$ are qualitatively different from those 
obtained in the simple QMC model. In particular, for $\delta$ = (13.0, 8.0, 3.6), 
corresponding to $g_\sigma^q$ = (0, 1.0, 2.0), we get $B/B_0 \simeq 0.36$ and
\begin{eqnarray}
M^*_N \simeq 660 - 720\, \mbox{MeV}\ ,
\label{tp-dirs}
\\*[7.2pt]
U_{\rm v}\simeq 150 - 215\, \mbox{MeV}\ ,
\label{tp-dirv}
\end{eqnarray}
at $\rho_N = \rho_N^0$. Since the equivalent scalar and vector 
potentials appearing in the wave equation for a point-like nucleon 
are $M_N^*-M_N$ and $U_{\rm v}$, respectively \cite{guichon95,blunden96}, 
Eqs.~(\ref{tp-dirs}) and (\ref{tp-dirv}) imply large and canceling scalar 
and vector potentials for the nucleon in nuclear matter.  Such potentials are
comparable to those suggested by Dirac phenomenology
\cite{hama90,wallace87}, Brueckner calculations \cite{wallace87}, and
finite-density QCD sum rules \cite{cohen95}, but smaller than those
obtained in QHD-I \cite{serot86}.  These potentials also imply a strong 
nucleon spin-orbit potential.  Therefore, the essential features of
relativistic nuclear phenomenology are recovered. The corresponding 
results for the nuclear matter compressibility, $K_V^{-1}$, are slightly 
larger than the corresponding value in the usual QMC model, but significantly 
smaller than that in QHD-I. The resulting total energy per nucleon for symmetric 
nuclear matter is shown in Fig.~\ref{fig-1}.

\begin{figure}[t]
\begin{center}
\epsfysize=11.7truecm
\leavevmode
\setbox\rotbox=\vbox{\epsfbox{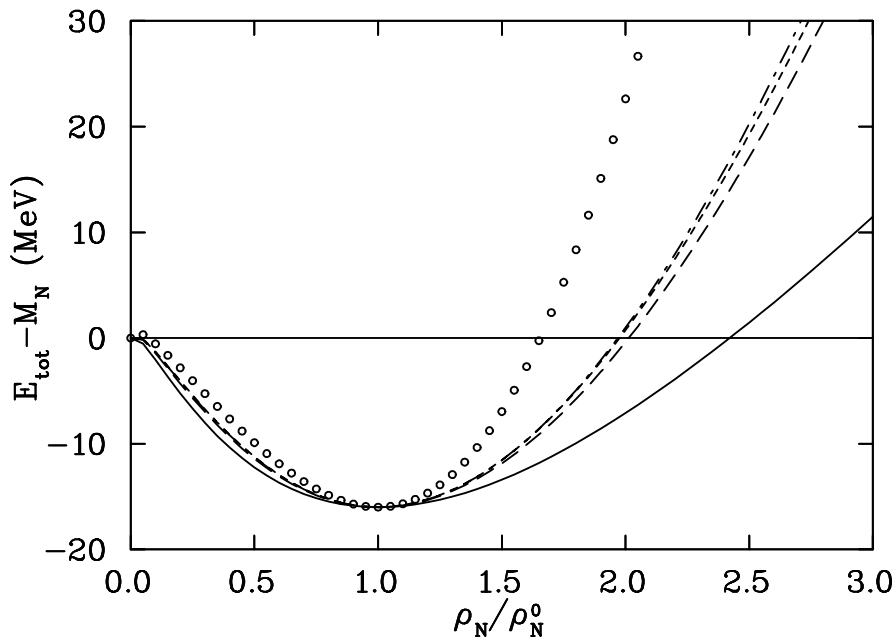}}\rotl\rotbox
\end{center}
\caption{Energy per nucleon for symmetric nuclear matter as a
function of the medium density, with $R_0=0.6$ fm and $\delta = 8$. 
Here the direct coupling model Eq.~(\protect{\ref{an-dir}}) for the in-medium 
bag constant is used. The solid curve corresponds to the usual QMC model, and 
the result from QHD-I is given by the open circles. The other three curves 
correspond to $g^q_\sigma =  0$ (long-dashed), 
1 (dot-dashed), and 2 (short-dashed), respectively.}
\label{fig-1}
\end{figure}
%

In the usual QMC model, the bag radius decreases slightly and the
quark root-mean-square (RMS) radius increases slightly in nuclear
matter with respect to their free-space values.  When the bag constant 
drops relative to its free-space value, the bag pressure decreases and 
hence the bag radius increases in the medium.  When the reduction of the
bag constant is significant, the bag radius in saturated nuclear matter 
is $25 - 30\%$ larger than its free-space value. The quark RMS radius 
also increases with density, with essentially the same rate as for the 
bag radius. This implies a ``swollen'' nucleon in nuclear medium, which 
has many attendant consequences
\cite{noble81,close83,celenza85,sick85,brown88,brown89,soyeur93}. It
is also interesting to note that the result of $25 - 30\%$ increase in
the nucleon size is comparable to those suggested in
Refs.~\cite{noble81,close83,celenza85} (see, however,
Ref.~\cite{sick85}).  In the special case, $g_\sigma = 0$ and
$\delta = 4$, we find that the reduction of $B$ in the nuclear matter
is too large ($B/B_0 \sim 10\%$). This leads to unreasonably large
values for the bag radius and the quark RMS radius in nuclear matter.

The results corresponding to the limit of $\delta\rightarrow \infty$
[i.e., Eq.~(\ref{dir-limit})] are also listed in
Table~\ref{tab-1}. These results are not far from those with finite
$\delta$ values. As $\delta$ increases, the results will approach
saturation. The last row of Table~\ref{tab-2} gives the results
obtained from fixing $g^q_\sigma$ at its value predicted by the simple
QMC model, 5.309 (for $R_0 = 0.6$ fm). In this case, it is found that 
for any given $\delta$ value the self-consistent solution requires 
$g^B_\sigma = 0$. When the coupling $g^q_\sigma$ is tuned from zero to its
corresponding value in the simple QMC model, the results interpolate
between the QHD-I results and the usual QMC model results. If
$g^q_\sigma$ exceeds its value in the simple QMC model, the in-medium
bag constant will increase instead of decrease relative to its
free-space value, which is in contradiction with the physics discussed
in the present paper.

For a fixed $g_\sigma^q$, the coupling constants $g_\sigma^B$ and
$g_\omega$ gets smaller as $\delta$ gets larger. Recall that the
reduction of $B$ is controlled by both $g_\sigma^B$ and
$\delta$. While $B/B_0$ and $M_N^*$ increase, $U_{\rm v}$, $R/R_0$,
and $K_V^{-1}$ decrease as $\delta$ increases. From Table~\ref{tab-1},
one can see that for a given value of $B/B_0$, one finds smaller $M_N^*$ and
larger $U_{\rm v}$ with a larger value of $g_\sigma^q$. We also find
that when $\delta$ gets too small, a self-consistent solution no
longer exists.

For curiosity, we have also explored the results for negative $\delta$
values. One can see from Eq.~(\ref{an-dir}) that negative $\delta$
values can also lead to a decreasing bag constant. In fact the
physical quantities are continuous as $1/\delta$ goes through zero. As
$1/\delta$ decreases from zero ($\delta$ goes negative), the scalar
and vector potentials continue the decrease seen in
Table~\ref{tab-1}. For $g^q_\sigma = 0$ and $\delta\sim -1.0$, the
resulting $M_N^*$ and $U_{\rm v}$ are similar to those found in the
usual QMC model. The compressibility is somewhat lower.

All of the above results use the direct coupling model; we now present 
the results for the scaling model. The two coupling constants,
$g_\sigma$ and $g_\omega$, are adjusted to fit the binding energy 
of saturated nuclear matter. There is only one free parameter,
$\kappa$, in this case. For various values of $\kappa$ and
free space bag radius $R_0$, the resulting coupling constants and
nuclear matter results are listed in Table \ref{tab-2}.

\begin{table} \caption{Coupling constants and nuclear matter results
as obtained from the scaling model. The case of $\kappa=0$ corresponds
to the simple QMC model and the last row gives the result of
QHD-I. Here the nuclear matter compressibility, $K_V^{-1}$, is given
in unit of MeV, $r_m^*$ and $r_m$ denote the quark root-mean-square
radius in nuclear matter and in free space, respectively, and $U_{\rm
v}\equiv g_\omega \overline{\omega}$ is the vector mean field. 
The mass parameters are the same as in Table~\protect{\ref{tab-1}}.}
\label{tab-2}
\begin{tabular}{cccccccccccc}
$R_0$(fm) &$\kappa$  &$g_\sigma^2/4\pi$  &$g_\omega^2/4\pi$ &$M_N^*/M_N$  
 &$U_{\rm v}/M_N$
&$K_V^{-1}$ &$B/B_0$ &$x/x_0$ &$R/R_0$ &$r_m^*/r_m$ \\
\tableline
\tableline
0.6            &0   & 20.18 & 1.56 & 0.89 & 0.04 & 223  & 1.0  & 0.93 & 0.98 
& 1.0  \\
               &1.0 & 11.90 & 2.27 & 0.87 & 0.06 & 258  &0.87  & 0.93 & 1.02 
& 1.03 \\
               &2.0 & 5.92  & 3.60 & 0.83 & 0.10 & 319  & 0.69 & 0.94 & 1.08 
& 1.10 \\
             & 2.95 & 2.24 & 7.78  & 0.71 & 0.22 & 590  & 0.36 & 0.94 & 1.27 
& 1.29 \\
               &3.0 & 2.11  & 8.32 & 0.69 & 0.24 & 628  & 0.33 & 0.94 & 1.30 
& 1.32 \\
\tableline
0.8            &0   & 22.01 & 1.14 &0.91 &0.03 & 202  & 1.0  & 0.90 & 0.99 & 
1.02 \\
               &1.0 & 12.78 & 1.76 &0.89 &0.05 & 235  & 0.89 & 0.91 & 1.02 & 
1.04 \\
               &2.0 & 6.20  & 2.88 &0.85 &0.08 & 289  & 0.73 & 0.93 & 1.08 & 
1.10 \\
               &3.0 & 2.06  & 6.39 &0.75 &0.18 & 479  & 0.42 & 0.93 & 1.24 & 
1.26\\
               &3.1 & 1.78  & 7.32 &0.72 &0.21 & 543  & 0.36 & 0.93 & 1.28 & 
1.31 \\
\tableline
1.0            &0   & 22.48 & 0.96 & 0.91 & 0.03& 192 & 1.0  & 0.88 & 1.0  & 
1.03 \\
               &1.0 & 12.94 & 1.54 & 0.89 & 0.04& 225 & 0.89 & 0.89 & 1.03 & 
1.05 \\
               &2.0 & 6.21  & 2.58 & 0.86 & 0.07& 276 & 0.75 & 0.91 & 1.07 & 
1.10 \\
               &3.0 & 2.01 & 5.68  & 0.77 & 0.16& 432 & 0.46 & 0.92 & 1.21 & 
1.24 \\
               &3.17 & 1.54 & 7.18 & 0.72 & 0.20 & 502 & 0.36 & 0.92 & 1.29 &
 1.32 \\
\tableline
QHD-I  &$\text{--}$ &8.45  & 12.84 & 0.55 & 0.37 &$540$  
& $\text{--}$ & $\text{--}$ & $\text{--}$ & $\text{--}$\\
\end{tabular}
\end{table}

Again, the decrease of the bag constant gives rise to the decrease of 
$M_N^*/M_N$ and the increase of $U_{\rm v}/M_N$ relative to their values
in the simple QMC model. We see from Table \ref{tab-2} that $M_N^*$
decreases and $U_{\rm v}$ increases rapidly as $\kappa$ increases.
For $\kappa = (2.95, 3.10, 3.17)$, corresponding to $R_0$ = (0.6, 0.8,
1.0 fm), we find $B/B_0 \simeq 0.36$ at the saturation density. The
corresponding results for $M_N^*$ and $U_{\rm v}$ are
\begin{eqnarray}
M_N^* &\simeq& 660 - 680\, \mbox{MeV} \ ,
\label{typical-s}
\\*[7.2pt]
U_{\rm v} &\simeq& 190 - 225\, \mbox{MeV} \ ,
\label{typical-v}
\end{eqnarray}
at $\rho_N = \rho_N^0$, which are similar to the direct coupling model 
results given in Eqs.~(\ref{tp-dirs}) and (\ref{tp-dirv}), though the
magnitudes for $M_N^* - M_N$ and $U_{\rm v}$ are slightly
larger. These results are also consistent with those suggested by
relativistic nuclear phenomenology and finite-density QCD sum
rules. 

One notices from Table~\ref{tab-2} that the value of $K_V^{-1}$
increases quickly when $\kappa$ value is increased. This results from
the increasing $g_\omega$ with increasing $\kappa$. For the
$\kappa$ values leading to Eqs.~(\ref{typical-s}) and
(\ref{typical-v}), the corresponding values for $K_V^{-1}$ are
comparable to that obtained in QHD-I, which is too large compared with
the empirical value.  This feature is also shown in Fig.~\ref{fig-2},
where the total energy per nucleon for symmetric nuclear matter is
plotted as a function of nuclear matter density for various $\kappa$
values, with $R_0=0.6$ fm. The result from QHD-I is also plotted for
comparison. The equation of state for the nuclear matter is much
softer in the simple QMC model than in QHD-I. As $\kappa$ gets larger,
the equation of state becomes stiffer.

\begin{figure}[t]
\begin{center}
\epsfysize=11.7truecm
\leavevmode
\setbox\rotbox=\vbox{\epsfbox{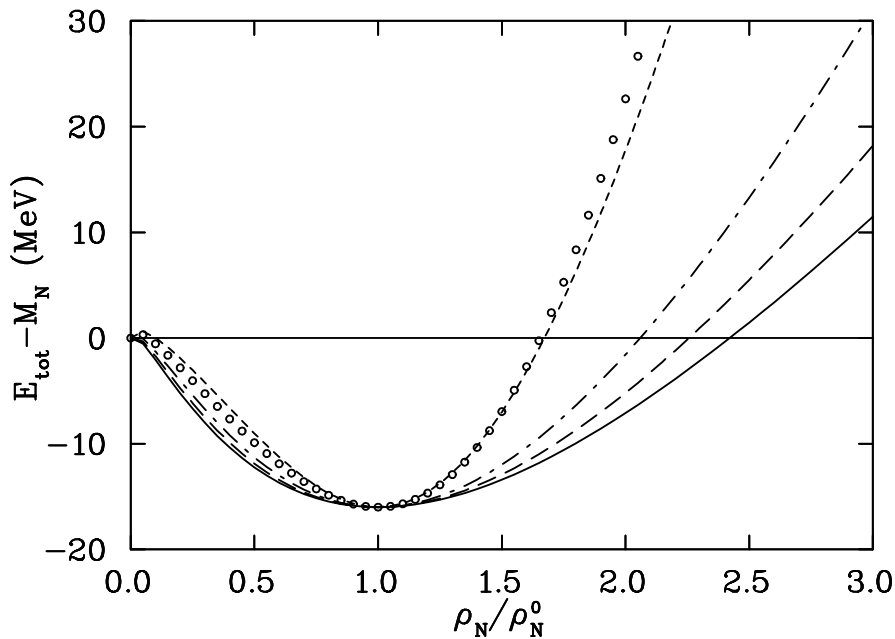}}\rotl\rotbox
\end{center}
\caption{Energy per nucleon for symmetric nuclear matter as a
function of the medium density, with $R_0=0.6$ fm. Here the scaling
model Eq.~(\protect{\ref{an-br}}) for the in-medium bag constant 
is adopted. The four curves correspond to $\kappa=0$ (solid), 1 (long-dashed), 
2 (dot-dashed), and 3 (short-dashed), respectively.
The result from QHD-I is given by the open circles.}
\label{fig-2}
\end{figure}

Both the bag radius and the quark RMS radius in nuclear matter are
larger than their free-space values. As discussed above, this is due
to the decrease of the bag constant in the nuclear medium. For the
$\kappa$ values yielding Eqs.~(\ref{typical-s}) and
(\ref{typical-v}), the corresponding bag radius and quark RMS radius at
$\rho_N = \rho_N^0$ are $25 -30\%$ larger than their free-space
value. This is essentially the same as that found in the direct coupling
model (see Table~\ref{tab-1}). 

The sensitivity of our results to the free-space bag radius $R_0$ is
also illustrated in Table~\ref{tab-2}. For a given $\kappa$ value, the
ratios $B/B_0$ and $M_N^*/M_N$ increase and the ratios $R/R_0$ and
$U_{\rm v}/M_N$ decrease as $R_0$ increases. However, for the $\kappa$
values considered here, the sensitivity of our results to $R_0$ is
small. The sensitivity of the results from the direct coupling model
to the choice of $R_0$ is also small and similar to that in the scaling 
model.

%
\section{Discussion}
\label{discussion}

As stressed by Saito and Thomas \cite{saito94}, in the simple QMC
model, all the effects of the internal quark structure of the nucleon
are summarized in the factor $C(\overline{\sigma})$. If
$C(\overline{\sigma}) = 1$ is a constant, one would get exactly the
same nuclear matter results as in QHD-I. In the simple QMC model,
$C(\overline{\sigma})$ is much smaller than unity, which leads to much
larger $M_N^*$ and much smaller $U_{\rm v}$ than those required in
QHD-I.

In the present study, we introduce the medium modification for the bag
constant.  As shown in Eqs.~(\ref{tc-d}) and (\ref{tc-k}), the effect
of this modification is completely absorbed into the factor
$C(\overline{\sigma})$.  We observe that for various parameters
considered here, both Eqs.~(\ref{tc-d}) and (\ref{tc-k}) lead to an
increase in $C(\overline{\sigma})$. This indicates that the reduction
of the bag constant in nuclear matter partially offsets the effect due
to the internal quark structure of the nucleon. It is thus not
surprising to find that our modified quark-meson coupling model gives
smaller $M_N^*$ and larger $U_{\rm v}$ than those found in the simple
QMC model. 

To further illustrate this point, we have plotted
$C(\overline{\sigma})$ in Fig.~\ref{fig-3} as a function of
$g^q_\sigma \overline{\sigma}$ for the usual QMC model and for our
modified QMC model with the scaling model for the in-medium bag
constant.  We see that in the simple QMC model, $C(\overline{\sigma})$
is small and decreases as $g^q_\sigma \overline{\sigma}$
increases. The introduction of dropping bag constant gives an increase
in $C(\overline{\sigma})$. When the reduction of bag constant is
large, $C(\overline{\sigma})$ is approximately constant and
significantly larger than one for small and moderate values of
$g^q_\sigma \overline{\sigma}$; as $g^q_\sigma \overline{\sigma}$
increases, $C(\overline{\sigma})$ decreases quickly. [The
self-consistent solution requires $g^q_\sigma \overline{\sigma} =$
(96, 82, 69, 58 MeV) at the saturation density, corresponding to
$\kappa =$ (0, 1.0, 2.0, 3.0), respectively.]  A similar plot can also
be done with the direct coupling model, but $C(\overline{\sigma})$ in
this case is a function of $g^q_\sigma$, $g^B_\sigma$, and
$\overline{\sigma}$, instead of $g^q_\sigma\, \overline{\sigma}$
alone.

\begin{figure}[t]
\begin{center}
\epsfysize=11.7truecm
\leavevmode
\setbox\rotbox=\vbox{\epsfbox{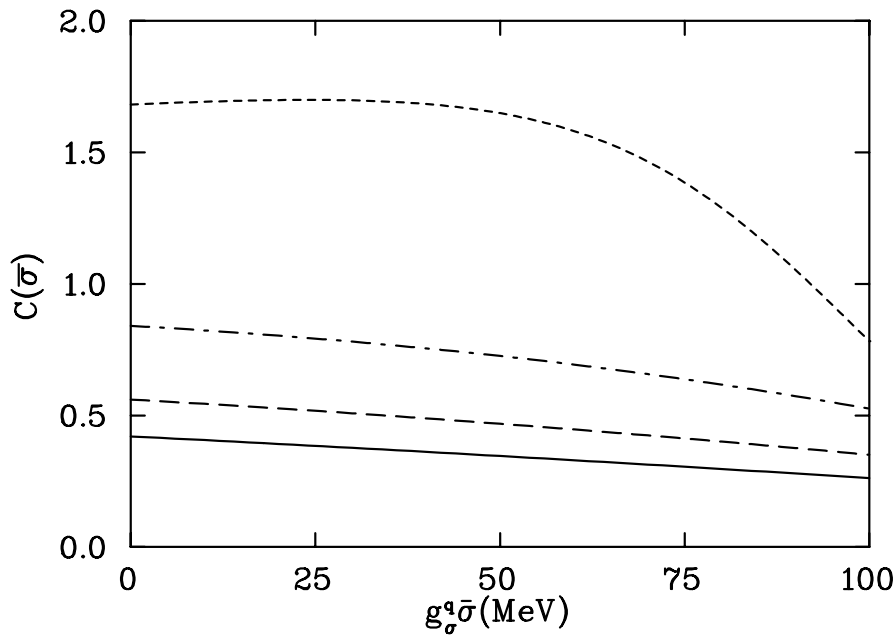}}\rotl\rotbox
\end{center}
\caption{The factor $C(\overline{\sigma})$ as a function of
$g^q_\sigma \overline{\sigma}$, with $R_0 = 0.6$ fm. The solid curve
is from the simple QMC model. The other three curves correspond to the
scaling model, with $\kappa = 1$ (long-dashed), 2 (dot-dashed), and 3
(short-dashed), respectively.}  
\label{fig-3} 
\end{figure} 

We observe that the extent to which the bag constant drops in
nuclear matter determines the physical outcome. 
Unless one expresses the bag constant in terms of QCD operators and
solves QCD in nuclear matter, the change of the bag constant in
nuclear medium is unknown. As such, one has to invoke model 
descriptions in order to obtain a quantitative estimate for the 
reduction of the bag constant in nuclear matter.

In Ref.~\cite{adami93}, Adami and Brown have argued that the MIT
bag constant is related to the energy associated with the chiral
symmetry restoration (the vacuum energy difference between the
chiral-symmetry-restored vacuum inside and the broken phase
outside). According to the scaling ansatz advocated by Brown and Rho
\cite{brown91}, the in-medium bag constant should scale like
\cite{adami93}, $B/B_0 \simeq \Phi^4$, where $\Phi$ denotes 
the universal scaling, $\Phi\simeq m_\rho^*/m_\rho \simeq f^*_\pi/f_\pi \cdots$,
which is density dependent. Here, the ``starred'' quantities refer to
the corresponding in-medium quantities. This scaling behavior is
argued to hold approximately at the mean-field level
\cite{adami93,brown91,brown95,brown95a}.  
Thus, one may get a rough estimate of the medium modification of the bag
constant from the medium modifications of vector-meson masses which
have been studied extensively
\cite{hatsuda92,jin95,kurasawa88,li95,hatsuda96}. Taking the result
for $m_\rho^*/m_\rho$ from the most recent finite-density QCD sum-rule
analysis \cite{jin95}, we find $\Phi\simeq m_\rho^*/m_\rho\sim 0.78$
at the saturation density, which gives rise to $B/B_0\simeq \Phi^4\sim
0.36$. This shows substantial reduction of the bag constant in
nuclear matter relative to its free-space value. With this estimate, we
obtain large and canceling scalar and vector potentials for the nucleon
in nuclear matter, which are consistent with those suggested by
relativistic nuclear phenomenology and finite-density QCD sum rules,
though smaller than those found in QHD-I. This feature is seen in both
models for the in-medium bag constant discussed here, implying a weak
model dependence of our results. 

However, some caveats concerning the above estimate must be added.
The Brown-Rho scaling is an ansatz based on the idea of partial chiral 
symmetry restoration in nuclear medium and the assumption that the scale
anomaly of QCD could be modeled by a light dilaton field \cite{brown91,adami93}.
It is unclear whether this ansatz can be justified in QCD. 
Although it has been argued that many nuclear phenomena are connected
to the partial chiral restoration in nuclear medium
\cite{brown88,brown89,soyeur93,adami93,brown91,brown95,brown95a,henley89,birse94},
the only compelling evidence for partial chiral symmetry
restoration in nuclear medium is that the magnitude of the chiral
quark condensate, $\langle\overline{q}q\rangle$,  is substantially 
reduced relative to its vacuum value \cite{drukarev90,cohen92,lutz92,birse93}. 
Recently, Birse \cite{birse96} has argued that the in-medium nucleon mass 
cannot be simply related to the change in the chiral quark condensate
and there are other important contributions unrelated to partial chiral
symmetry restoration. Moreover, it is known that both the MIT bag model and the 
QHD are not compatible with the chiral symmetry. 

Clearly, how the bag constant changes in nuclear matter is an important 
topic for further study. The investigation of finite nuclei and nuclear 
structure functions in the present model may offer some independent 
information and/or constraints on the modification of the bag constant. 
Work along this direction is in progress \cite{jin96a}. Another direction 
may be in motivating an explicit functional form for the in-medium bag
constant, $B(\overline{\sigma})$, from a soliton-like model for the nucleon.
Recently, an explicit form for $B(\overline{\sigma})$ has been 
suggested in Ref.~\cite{cahill96}, based on the global color model 
of QCD.

The quark-meson coupling model for nuclear matter is valid only if the
nucleon bags do not overlap significantly. In his original
paper\cite{guichon88}, Guichon has suggested $R_0=0.6-0.7$ fm in order
to keep the overlapping effect small.  In our modified QMC model, both
the bag radius and the quark RMS radius increase in medium relative to
their values in free-space due to the dropping bag constant in
medium. In Fig.~\ref{fig-4}, the spatial quark wave functions are plotted 
as functions of radial coordinate with the scaling model for the in-medium
bag constant. We see that when the bag constant drops the quark wave 
functions are pushed outward. This depicts a ``swollen'' nucleon picture, 
which has important implications in many nuclear physics issues
\cite{noble81,close83,celenza85,sick85,brown88,brown89,soyeur93}.

\begin{figure}[t]
\begin{center}
\epsfysize=11.7truecm
\leavevmode
\setbox\rotbox=\vbox{\epsfbox{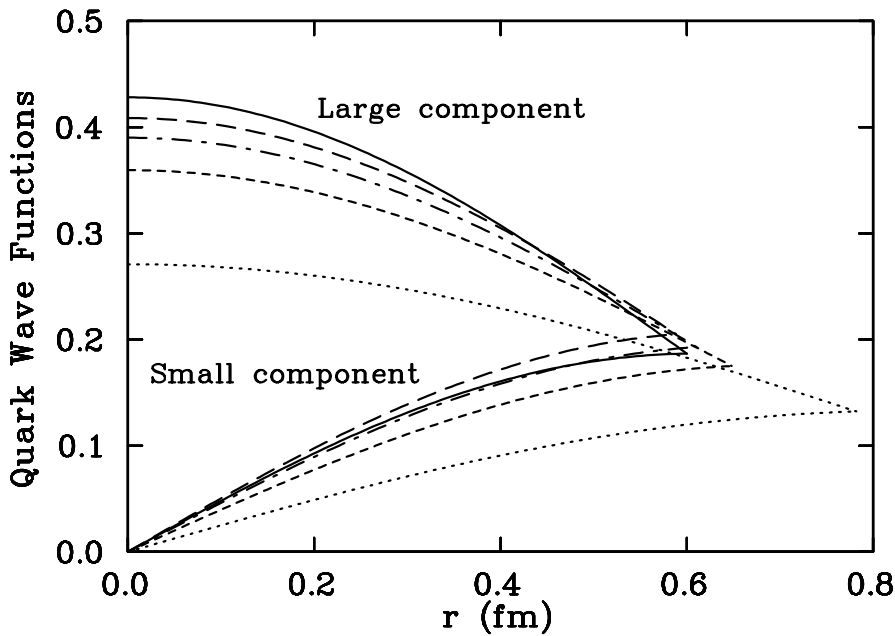}}\rotl\rotbox
\end{center}
\caption{Quark wave functions as functions of radial coordinate. Here the scaling
model Eq.~(\protect{\ref{an-br}}) for the in-medium bag constant 
is adopted. The solid curves are the free-space quark wave functions. 
The other four set of curves correspond to, $\kappa=0$ (long-dashed), 
1 (dot-dashed), 2 (short-dashed), and 3 (dotted), respectively.}
\label{fig-4}
\end{figure}
  
On the other hand, the increasing bag radius also implies larger
overlapping effect than in the usual QMC model.  When the bag constant
in nuclear matter is significantly smaller than its free-space value
($B/B_0 \sim 0.4$), we find $R/R_0 \sim 1.25 - 1.30$ at the saturation
density, which gives $4 \pi R^3 \rho^0_N/3\sim 0.3 - 0.34$ when $R_0 =
0.6$ fm. This indicates that the overlap between the bags is still
reasonably small at the saturation density, though a factor of two
larger than in the usual QMC model. For larger $R_0$ and/or higher
densities, the overlap becomes more significant and the
non-overlapping bag picture of the nuclear matter may become
inadequate. However, it is unclear at this stage whether the overlap
between the bags is already included effectively in the reduction of
the bag constant and/or in the scalar and vector mean fields. Further
study is needed to clarify such issue.  We also note that in the
scaling model with large $\kappa$ values, the resulting nuclear matter
compressibility is too large compared to the empirical value.  This
may be fixed by introducing self-interactions of the scalar field,
which, however, will introduce more free parameters. The direct
coupling model, on the other hand, produces a more reasonable value
for nuclear matter compressibility.

The QMC model is probably the simplest extension of QHD to incorporate
explicit quark degrees of freedom, where the exchanging mesons are
treated as classical fields in the mean-field approximation. To be
more consistent, the explicit quark structures of the mesons should
also be included and the physics beyond the mean-field approximation
should be considered. It has been emphasized above that both the
MIT bag model and the QHD are not chirally symmetric. 
As such, the quark-meson coupling models discussed here are not chiral
models. At the hadron level, significant progress has been made in 
incorporating both chiral symmetry and broken scale invariance in 
relativistic hadronic models \cite{heide94,furnstahl93}. So, the 
quark-meson coupling models may be extended by combining these hadronic 
chiral models and a chiral version of the bag model. Recently, it has also
been argued that connections can be made between effective chiral
Lagrangians and QHD model at the mean-field level \cite{brown96}.

%
\section{Summary} \label{conclusion} 

In this paper, we have modified the quark-meson coupling model by 
introducing medium modification of the bag constant. We proposed 
two models for the in-medium bag constant, direct coupling model
and scaling model. The former couples the bag constant directly to 
the scalar mean field, and the latter uses a scaling ansatz which 
relates the in-medium bag constant to in-medium nucleon mass. Both 
models feature a decreasing bag constant with increasing density.

The reduction of the bag constant in nuclear matter partially offsets
the effect of the internal quark structure of the nucleon and effectively 
introduces a new source of attraction. This attraction needs to be 
compensated with additional vector field strength. The decrease of 
bag constant also implies the increase of bag radius in nuclear matter. 
This is consistent with the ``swollen'' nucleon picture discussed in the
literature.

When the bag constant is reduced significantly in nuclear matter with
respect to its free-space value, we find that our modified quark-meson 
coupling model predicts large and canceling scalar and vector potentials 
for the nucleon in nuclear matter, which is qualitatively different 
from the prediction of the simple QMC model. These potentials are 
consistent with those suggested by the relativistic nuclear phenomenology 
and finite-density QCD sum rules. The internal quark structure of the 
nucleon seems to play only a relatively minor role. On the other hand, 
the reduction of the bag constant in nuclear medium relative to its 
free-space value may play an important role in low- and medium-energy
nuclear physics.

\acknowledgments
This work was supported by the Natural Sciences and Engineering 
Research Council of Canada.

\appendix
\label{app}
\section*{A}

In this appendix, we demonstrate that adopting Eq.~(\ref{an-dir}) with
$\delta = 4$ for the in-medium bag constant and taking $g^q_\sigma =
0$ (and $m^0_q = 0$), one can reproduce the QHD-I results for nuclear
matter. To this end, we show that the resulting expressions for the
total energy per nucleon and the self-consistency condition for the
scalar mean field are identical to those in QHD-I.

In free space, the nucleon mass and the equilibrium condition for 
the bag are given by
\begin{eqnarray}
& &M_N=\sqrt{E_{\rm bag}^2-3 x_0^2/R_0^2}\ ,
\\*[7.2pt]
& &E_{\rm bag}\left[ -3{x_0\over R_0^2}+ {Z_0\over R_0^2} +
4 \pi R_0^2 B_0\right]
+3 {x_0^2\over R^3_0} = 0\ ,
\label{vac-set}
\end{eqnarray}
where 
\begin{equation}
E_{\rm bag}=3 {x_0\over R_0} - {Z_0\over R_0 } +{4\over 3} \pi R_0^3 B_0\ .
\label{aebag}
\end{equation}
Solving these two equations, one finds that the combinations $B_0 R_0^4$ 
and $M_N R_0$ can be expressed in terms of $x_0$ and $Z_0$. This can also be 
seen easily from dimensional considerations. Since these two combinations
are dimensionless, they must depend only on the dimensionless parameters
$x_0$ and $Z_0$.

In the nuclear medium, the corresponding two equations become
\begin{eqnarray}
& &M_N^*=\sqrt{E_{\rm bag}^2-3 x^2/R^2}\ ,
\\*[7.2pt]
& &E_{\rm bag}\left[ -3{x\over R^2}+ {Z\over R^2} +4 \pi R^2 B\right]
+3 {x^2\over R^3} = 0\ .
\label{med-set}
\end{eqnarray}
Similarly,  $B R^4$ and $M_N^* R$ are only dependent on $x$ and $Z$. 
Since we take $g^q_\sigma = 0$ and $Z=Z_0$, 
$x (= x_0)$ and $Z$ are independent of the nuclear matter density.
One therefore concludes that
\begin{equation}
{B\over B_0} = \left( {M_N^*\over M_N}\right )^4 = 
\left ( {R_0\over R} \right)^4\ .
\label{sca-rel}
\end{equation}
Using Eq.~(\ref{an-dir}) with $\delta= 4$, one gets
\begin{equation}
{M_N^*\over M_N} = 1- {g^B_\sigma \overline{\sigma} \over M_N}\ .
\end{equation}
One can then rewrite the total energy per nucleon at finite nuclear 
matter density as
\begin{eqnarray}
E_{\rm tot}& = &{\gamma\over (2\pi)^3\, \rho_N} 
\int^{k_F}\, d^3 k \sqrt{M_N^{* 2}+{\bf k}^2}
+{g_\omega^2\over 2 m_\omega^2}\,\rho_N
+{m_\sigma^2\over 2\,(g^B_\sigma)^2\,\rho_N}(M_N-M_N^*)^2\ .
\end{eqnarray}
The self-consistency condition for the scalar field, Eq.~(\ref{scc}), 
can be expressed as
\begin{equation}
M_N^* = M_N-{(g^B_\sigma)^2\over m_\sigma^2}\, 
{\gamma\over (2\pi)^3} 
\int^{k_F}\, d^3 k\, {M_N^*\over 
\sqrt{M_N^{* 2}+{\bf k}^2}} \ .
\end{equation}
These two equations are identical to those required in QHD-I. 
Thus, fitting the nuclear matter binding energy at the saturation
density, one should find the same scalar and vector couplings
and hence the same strengths for scalar and vector mean fields
as obtained in QHD-I.



\end{document}